\documentclass[runningheads]{llncs}
\usepackage{graphicx}
\usepackage{comment}

\begin{document}
%
%

\title{Phenotype Search Trajectory Networks for Linear Genetic Programming}
%
%

\author{Ting Hu\inst{1}\orcidID{0000-0001-6382-0602} \and
Gabriela Ochoa\inst{2}\orcidID{0000-0001-7649-5669} \and
Wolfgang Banzhaf\inst{3}\orcidID{0000-0002-6382-3245}}

\authorrunning{T. Hu et al.}
%
\institute{School of Computing, Queen's University, Kingston, ON K7L 2N8, Canada
\email{ting.hu@queensu.ca}\\
\and
University of Stirling, Stirling, FK9 4LA, UK\\
\email{gabriela.ochoa@stir.ac.uk}%
\and%
Department of Computer Science and Engineering, BEACON Center for the Study of Evolution in Action, and Ecology, Evolution and Behavior Program, Michigan State University, East Lansing, MI 48864, USA\\
\email{banzhafw@msu.edu}}

\maketitle              
\begin{abstract}
Genotype-to-phenotype mappings translate genotypic variations 
such as mutations into phenotypic changes.
Neutrality is the observation that some mutations do not lead to phenotypic changes.
Studying the search trajectories in genotypic and phenotypic spaces,
especially through neutral mutations,
helps us to better understand the progression of evolution and its algorithmic behaviour.
In this study,
we visualise the search trajectories of a genetic programming system as graph-based models,
where nodes are genotypes/phenotypes and edges represent their mutational transitions.
We also quantitatively measure the characteristics of phenotypes
including their genotypic abundance (the requirement for neutrality) and Kolmogorov complexity.
We connect these quantified metrics with search trajectory visualisations,
and find that more complex phenotypes are under-represented by fewer genotypes
and are harder for evolution to discover.
Less complex phenotypes, on the other hand,
are over-represented by genotypes, 
are easier to find, 
and frequently serve as stepping-stones for evolution.

\end{abstract}
\setcounter{footnote}{0}

\section{Introduction}

Neutral networks have been found to play an important role 
in natural and artificial evolution~\cite{banzhaf2006,reidys1997,shipman2000}. 
The notion of neutral networks derives from the idea that 
a search space can be explored by neutral moves 
as well as by moves improving fitness, 
with nodes of such a network the genotypes being visited 
and edges between them the variation steps taken by a searcher on that network. 
Each node, being a genotype also carries a fitness which can be used to determine 
whether a move from another node is allowed or not. 
Some researchers have claimed that neutral moves are extremely important 
to allow evolutionary progress to proceed~\cite{kimura1983}, 
and our long-standing interest and understanding of the role of neutrality 
in genetic programming (GP) systems
is deepened by the examination we report here. 

\begin{figure}[t]
\begin{center}
\includegraphics[width=.7\textwidth]{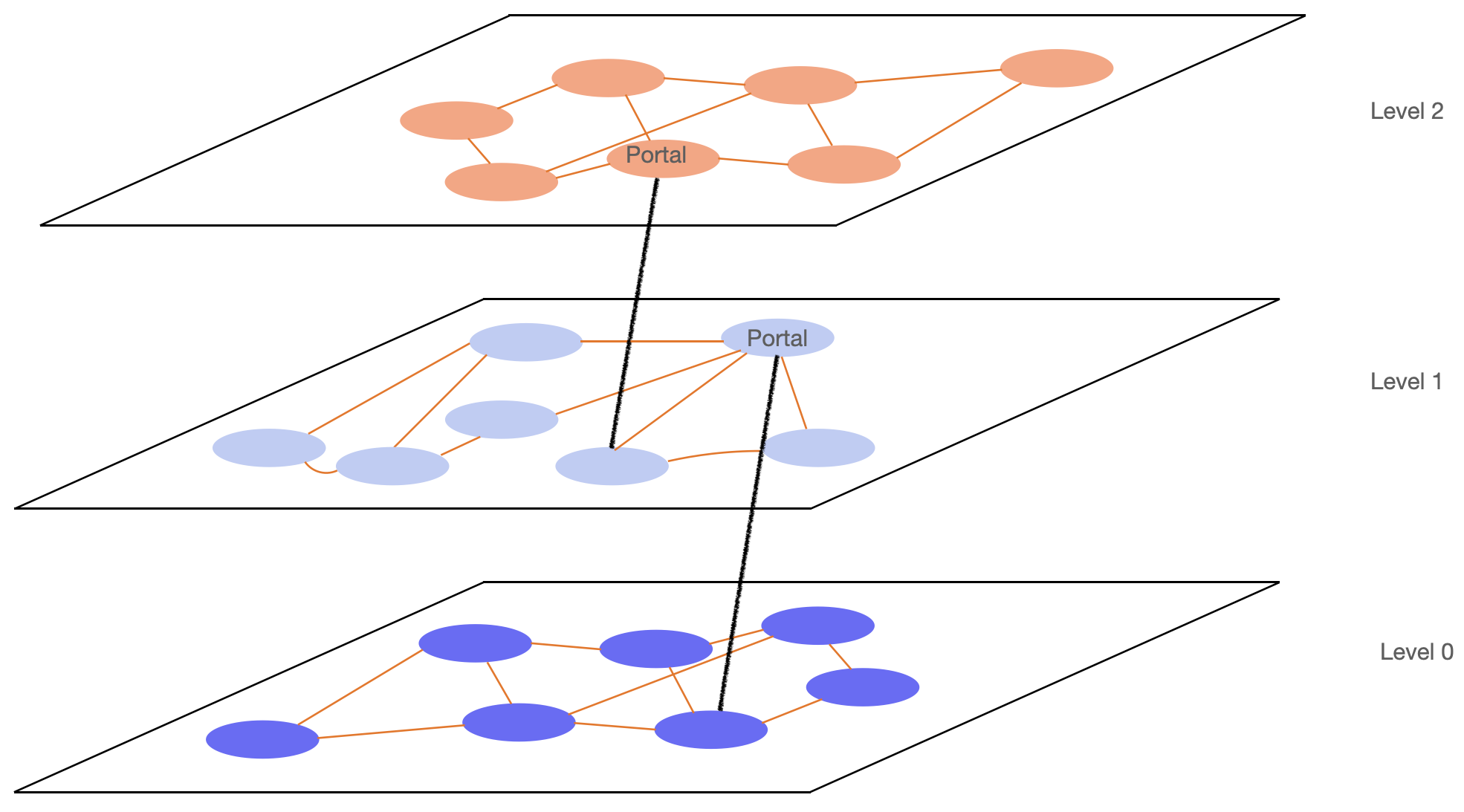}
\caption{Sketch of a network of neutral networks. 
Each level depicts one neutral network, with a discrete fitness value corresponding to its level. 
Nodes depict genotypes (genetic programs) which are connected within a level, 
reachable by neutral moves, with few nodes allowing to jump to a lower level (better fitness). 
The fitness of a node is measured by executing it 
and comparing the function it stands for with a target relation. 
The neutral networks are connected through what are called {\it portal} nodes 
to other neutral networks at a lower (better) fitness level.} \label{fig:n-o-n-sketch}
\end{center}
\end{figure}

Figure~\ref{fig:n-o-n-sketch} shows a sketch of 
how we conceptualize the search space of a GP algorithm: 
As a network of networks~\cite{gao2022,gao2014}. 
A single searcher in such a search process starts likely at a less-fit level (fitness level 2), 
and moves through the network by hopping from node to node via mutations or other variation operations. 
Occasionally, a {\it portal} node is found this way, 
which allows the searcher to enter a better fitness level. 
At that point, search again moves through the network until it finds another portal.

Studying such search trajectories and connect them with quantified metrics of genotypes and phenotypes
allow us to better understand the genotype-to-phenotype maps (G-P maps) 
and search behaviours of evolutionary algorithms.
In this research, we adapt a recent graph-based model, search trajectory networks (STNs)~\cite{stn_evostar,stn_main}
to analyse and visualise search trajectories of a simple linear GP system used to evolve Boolean functions. 
Search trajectory networks are a data-driven, graph-based model of search dynamics 
where nodes represent a given state of the search process 
and edges represent search progression between consecutive states. 
We connect this visualisation with an examination of the statistical behavior of 
those searchers navigating the corresponding genotype space. 
Following more recently formulated ideas about stiff G-P maps, 
we can tie the complexity of phenotypes to their potential for serving as stepping stones 
to a solution of the problem. 

We define stiff G-P maps as those maps that have a strong correlation 
between the complexity of the genotype and the complexity of the corresponding phenotype. 
In nature, such maps can be found in the molecular world, 
and in computing they are found in the conventional maps of GP. 
However, there is a different kind of maps 
that allows the correlation between genotype complexity and phenotype complexity to relax, 
typically found in developmental systems, 
which allow the complexity of the phenotype to grow over time under the influence of a genotype. 
The results reported here might not apply to such a kind of maps, 
though they are certainly realizable in GP.

Our research provides novel insights into how G-P maps result in
the heterogeneity of phenotypes being represented by genotypes,
i.e., some phenotypes are over-represented by a large number of genotypes (high redundancy)
but some are under-represented by few (low redundancy).
We find the correlation between the redundancy of a phenotype and its complexity
and show how they influence search trajectories and progression.

\section{The LGP System}

\subsection{Boolean LGP algorithm}

The GP algorithm used in our research is a linear genetic programming (LGP) system
where a sequential representation of computer programs is employed 
to encode an evolutionary individual~\cite{brameier2007}.
Such a linear genetic program is often comprised of a set of imperative instructions
to be executed sequentially.
Registers are used to either read input variables (input registers) 
or to enable computational capacity (calculation register).
One or more registers can be designated as the output register(s)
such that the final stored value(s) after the program is executed 
will be the program's output.

In this study,
we use an LGP algorithm for a three-input, one-output Boolean function search application.
Each instruction has one return, two operands and one Boolean operator.
The operator set has four Boolean functions \{{\tt AND}, {\tt OR}, {\tt NAND}, {\tt NOR}\},
any of which can be selected as the operator for an instruction.
Three registers ${\tt R_1}$, ${\tt R_2}$, and ${\tt R_3}$ receive the three Boolean inputs,
and are write-protected in a linear genetic program.
That is, they can only be used as an operand in an instruction.
Registers ${\tt R_0}$ and ${\tt R_4}$ are calculation registers,
and can be used as either a return or an operand.
Register ${\tt R_0}$ is also the designated output register,
and the Boolean value stored in ${\tt R_0}$ after a linear genetic program's execution
will be the final output of the program.
All calculation registers are initialized as {\tt FALSE} before execution of a program.
An example linear genetic program with three instructions is given as follows.
\begin{eqnarray*}
I_1: {\tt R_4 = R_2} & \texttt{AND} & {\tt R_3}\nonumber \\
I_2: {\tt R_0 = R_1} & \texttt{OR} & {\tt R_4}\nonumber \\
I_3: {\tt R_0 = R_3} & \texttt{AND} & {\tt R_0}\nonumber
\end{eqnarray*} 

A linear genetic program can have any number of instructions,
however, for the ease of sampling in this study,
we determine that a linear genetic program has a fixed length of six instructions.

\subsection{Genotype, phenotype, and fitness}

The {\it genotype} in our GP algorithm is a unique linear genetic program.
Since we have a finite set of registers and operators, as well as a fixed length for all programs,
the genotype space is finite.
Specifically, considering an instruction,
two registers can be chosen as the return, all five registers can be used as the two operands,
and the operator is picked from the set of four possible Boolean functions.
Thus, there are $2\times 5 \times 5 \times4 = 200$ unique instructions.
Given the fixed length of six instructions for all linear genetic programs,
we have a total number of $200^6 = 6.4\times 10^{13}$ possible different programs.

The {\it phenotype} in our GP algorithm is a Boolean relationship 
that maps three inputs to one output, represented by a linear genetic program,
i.e., $f: {\mathbf B}^3 \rightarrow {\mathbf B} $, 
where ${\mathbf B}$ = \{{\tt TRUE}, {\tt FALSE}\}.
There are thus a total of $2^{2^3} = 256$ possible Boolean relationships.
Having $6.4\times 10^{13}$ genotypes to encode 256 phenotypes,
our LGP algorithm must have a highly redundant genotype-phenotype mapping.
We define the {\it redundancy} of a phenotype
as the total number of genotypes that map to it.

The {\it fitness} of a linear genetic program 
is dependent on the target Boolean relationship (phenotype),
and it is defined as the dissimilarity of the calculated and the target Boolean relationships.
Given three inputs, there are $2^3=8$ combinations of Boolean inputs.
The Boolean relationship encoded by a linear genetic program can be seen as an 8-bit string 
representing the outputs that correspond to all 8 possible combinations of inputs.
Fitness is defined as the Hamming distance of this 8-bit output and the target output.
For instance, 
if the target relationship is 
$f({\tt R_1},{\tt R_2},{\tt R_3}) = {\tt R_1}\;\texttt{AND}\;{\tt R_2}\;\texttt{AND}\;{\tt R_3}$,
represented by the 8-bit output string of 00000001,
the fitness of a program encoding the FALSE relationship, 
i.e., 00000000, is 1.
Fitness is to be minimized and falls into the range between 0 and 8, 
where 0 is the perfect fitness and 8 is the worst.

\section{Kolmogorov Complexity}
\label{sec:complexity}

Dingle et al.~\cite{dingle2018} report a very general result 
on complexity limited discrete input-output maps. 
Based on algorithmic information theory 
they state that the probability of finding certain outputs 
depends on their Kolmogorov complexity. 
In particular, the probability to find an output $x \in O$ can be bounded 
by a quantity that depends exponentially on its Kolmogorov complexity:
\begin{equation} \label{eq:prob}
    P(x) \leq 2^{-(K(x|f,n)+\mathcal{O}(1))},
\end{equation}
where $K(x|f,n)$ is the shortest program that produces $x$, given $f$ and $n$, 
where $f$ is the computable input-output map 
$f:I \rightarrow O$ and $n$ characterizes the size of the input space. 
For binary inputs their number would be $2^n$. 
While this gives only an upper bound, 
we can see that it is (negatively) exponentially dependent on complexity, 
and if one compares two outputs, 
this fact can be used to predict the prevalence of one output over the other. 
Even more astonishing, this estimate becomes independent of the particulars of the map itself, 
in the above mentioned asymptotic case of a limited complexity map 
where $K(f) + K(n) << K(x) + \mathcal{O}(1)$:
\begin{equation}
    K(x|f,n) \approx K(x) + \mathcal{O}(1).
\end{equation}
In a later paper Dingle et al.~\cite{dingle2020} apply these findings to a variety of systems, 
among them the RNA G-P map (from linear sequence to 2D structure) and to others.   

Here we shall use these ideas to explain and predict the phenotypic trajectories of adaptive walkers 
in the fitness landscape of Boolean functions~\cite{vanneschi2012}. 
In the context of our LGP algorithm,
we define the Kolmogorov complexity (K-complexity) of a phenotype (Boolean relationship)
as the minimal {\it effective} length of its underlying linear genetic programs.
The effective length of a linear genetic program
is the number of its effective instructions.
An instruction of a program is effective
when its execution influences the final result of the output, here the content of register ${\tt R_0}$.
We can then conceptualize the search process 
as an adaptive walk in the network of solutions (phenotypes), 
and, by repeating the process with a number of runs, 
we can visualise the prevalence of certain transitions (hops of searchers in the network). 


\section{Sampling and Metrics Estimation}
\label{sec:sampling}

Although finite,
the genotype space of our LGP algorithm is enormous with a size of $6.4\times 10^{13}$ and 
can be challenging for exhaustive enumeration.
Therefore, we conduct sampling 
by randomly generating one billion linear genetic programs 
($\approx 0.00156\%$ of the total possible programs)
to approximate the genotype space.

We then map these one billion programs 
to the Boolean relationships (phenotypes) they represent,
and estimate the redundancy of each phenotype
as the total number of sampled genotypes that map to it.
239 out of the 256 phenotypes are represented by our sampled genotypes,
among which phenotype FALSE has the greatest redundancy of almost 109 million genotypes,
i.e., $>1\%$ of the total number of sampled genotypes.

\begin{figure}[t]
\begin{center}
\includegraphics[width=\textwidth]{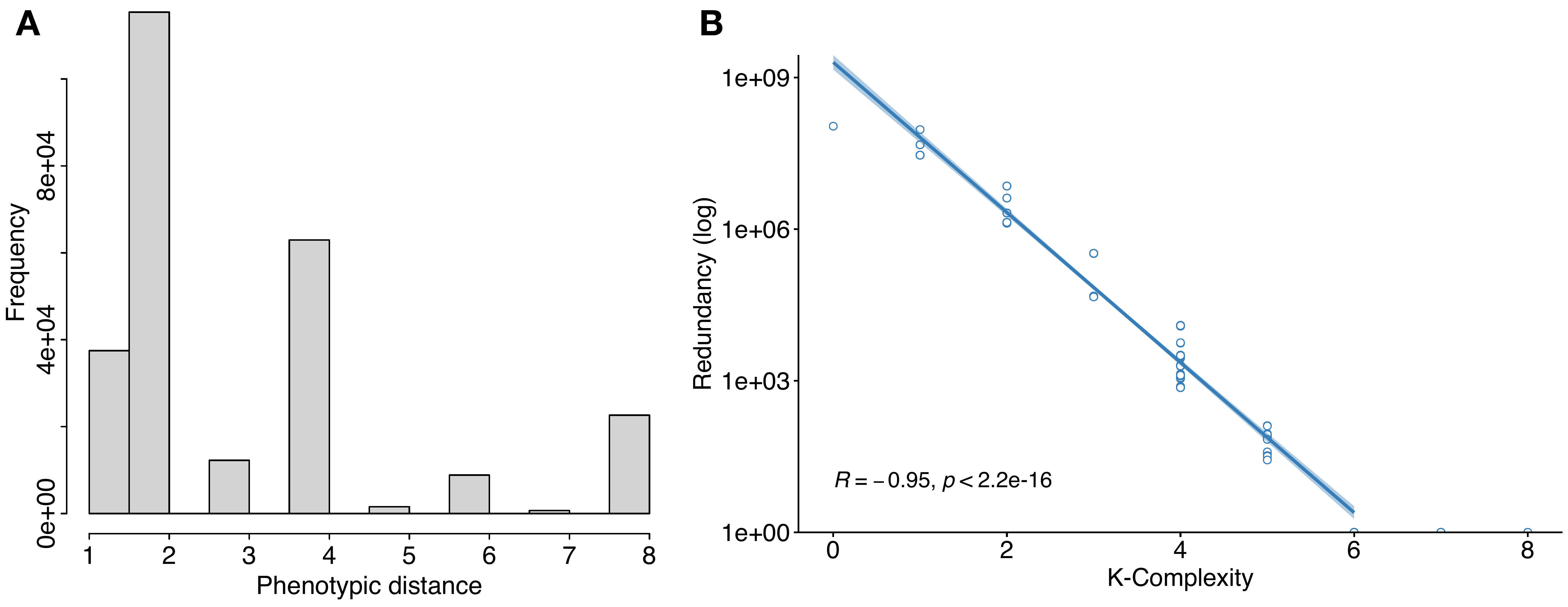}
\caption{Estimated metrics and characteristics.
(A) Distribution of phenotypic distance of one-step genotype neighbours. 
Neutral mutations ($73.8\%$ of all mutations) are excluded from the graph.
(B) Correlation of redundancy and K-complexity in log-linear scale.
Phenotypes are represented as circles. 
Note the log scale; the straight line is the best fit to an exponential decay.
Spearman's rank correlation coefficient $R$ and $p$-value are also provided.}  
\label{fig:redun_complexity}
\end{center}
\end{figure}

We first investigate the phenotypic effects of point mutations in our LGP system
by sampling one million genotypes and their one-step mutants.
Given the high redundancy in the G-P map of this system,
we observe that about $73.8\%$ of the sampled point mutations are neutral.
For the $26.2\%$ non-neutral mutations,
we compute the phenotypes of the genotype pairs for each mutation
and measure the Hamming distance of these phenotypes.
Figure~\ref{fig:redun_complexity}A shows the distribution of such pairwise phenotypic distances.
We see that the majority of non-neutral point mutations results in small phenotypic changes 
but also that there is a substantial number of mutations with larger step sizes (4 or even 8 bits).

Next, recall that the K-complexity of a phenotype is the minimal effective length of its underlying programs.
Essentially, the goal is to search the shortest effective program 
that can encode a given Boolean function (phenotype).
Again we randomly sample one billion linear genetic programs with varying lengths
drawn from the range between 5 and 20.
We then perform the structural intron removal algorithm~\cite{brameier2007}
to identify the effective length of each program.
We record for each phenotype
the minimal effective length of its sampled underlying programs,
and use it to estimate the K-complexity of that phenotype.
Figure~\ref{fig:redun_complexity}B shows the correlation of
redundancy and K-complexity for all the phenotypes we sampled and measured.
A strong and significant negative correlation is observed,
which means that more complex phenotypes are represented by fewer genotypes, 
as suggested by~\cite{dingle2018}.

To study the search trajectories for our LGP system,
we perform adaptive walks where only neutral or improving point mutations are accepted.
For a comparison, 
we set three target phenotypes with increasing difficulties,
i.e., an easy target of phenotype 240 (redundancy 46 million, K-complexity 1),
a medium target 20 (redundancy 3130, K-complexity 4),
and a hard target 30 (redundancy 772, K-complexity 4).
Two search scenarios are implemented,
where first we always start with a randomly generated genotype 
of the most distant phenotype from the target, i.e., fitness of 8,
and second we randomly generate a genotype without any consideration on its fitness.
We call the first scenario {\it fixed start} search and the second {\it random start} search.
We collect 100 runs for each scenario with each target phenotype,
where in each run we initialize a linear genetic program 
and let it walk in the search space for 2,000 steps.
These results are used for the visualisation of the search trajectories.

\section{Search Trajectory Networks}
\label{sec:stns}

Search trajectory networks (STNs) \cite{stn_evostar,stn_main} are a graph-based tool 
to visualise and analyse the dynamics of any type of meta-heuristic: 
evolutionary, swarm-based or single-point, on both continuous and discrete search spaces. 
Originally, the model tracks the trajectories of search algorithms in genotypic space, 
where nodes represent visited genotypes. 
However, for very large search spaces, 
techniques have been proposed to cluster sets of genotypes into \emph{locations}~\cite{stn_main} 
which can even group genotypes with the same phenotype or behavior~\cite{SartiAO22}, 
in order to have coarser models that can be visualised and interpreted.

In order to define a graph-based model, 
we need to specify its nodes and edges. 
We start by giving these general definitions before 
describing three STN models we propose here to visualise GP search spaces.

\subsection{General definitions}
\begin{description}
\item{\emph{Representative solution}.} 
A solution (genotype) to the optimization problem at a given time step 
that represents the status of the search algorithm 
(e.g.~best in the population in a given iteration, 
incumbent solution for single point meta-heuristics).

\item{\emph{Location}.} 
A non-empty subset of solutions 
that results from a predefined coarsening of the search space.

\item{\emph{Trajectory}.}  
Given a sequence of representative solutions in the order in which 
they are encountered during the search process, 
a search trajectory is defined as a sequence of locations formed 
by replacing each solution with its corresponding location.

\item{\emph{Nodes} ($N$).} 
The set of locations in a search trajectory of the search process being modeled. 

\item{\emph{Edges} ($E$).} 
Directed, connecting two consecutive nodes in the search trajectory. 
Edges are weighted with the number of times a transition between two given nodes occurred 
during the process of sampling and constructing the STN.

\item{\emph{STN}.} 
Directed graph STN = ($N$, $E$), with nodes $N$ and edges $E$ as defined above.
\end{description}

\subsection{The proposed STN models}

We propose three models with increasing coarsening, 
that is, with nodes grouping an increasing number of candidate solutions,  
in order to visualise the large and extremely neutral LGP search space under study.
\begin{enumerate}
\item{\bf Genotype STN}. 
The locations (nodes) are unique genotypes in the search space, 
and edges represent transitions between genotypes.	
\item{\bf Genotype-Phenotype STN}. 
The locations (nodes) are phenotypes grouping connected components in the Genotype STN 
that share the same phenotype. 
Edges represent transitions between (compressed) nodes.	
\item{\bf Phenotype STN}. 
The locations (nodes) are unique phenotypes in the search space, 
and edges represent consecutive transitions between phenotypes.	
\end{enumerate}

For constructing the STN models, 
multiple runs of adaptive walks (described in Section~\ref{sec:sampling}) are performed, 
and the visited locations and their transitions are aggregated into a single graph model. 
Notice that some locations and transitions may appear multiple times during the sampling process. 
However, the graph model retains as nodes each unique location, 
and as edges each unique transition between visited locations. 
Counters are maintained as attributes of the graph, 
indicating the frequency of occurrence of each (unique) node and edge. 

\subsection{Network visualisation}

Visualisation is a powerful and aesthetically inspiring way of appreciating network structure, 
which can offer insights not easily captured by network metrics alone. 
Node-edge diagrams are the most familiar form of network visualisation, 
where nodes are assigned to points in the two-dimensional Euclidean space 
and edges connect adjacent nodes by lines or curves. 
Nodes and edges can be decorated with visual properties such as size, 
color and shape to highlight relevant characteristics.

\begin{figure}[t]
\centering
\includegraphics[width=.95\linewidth]{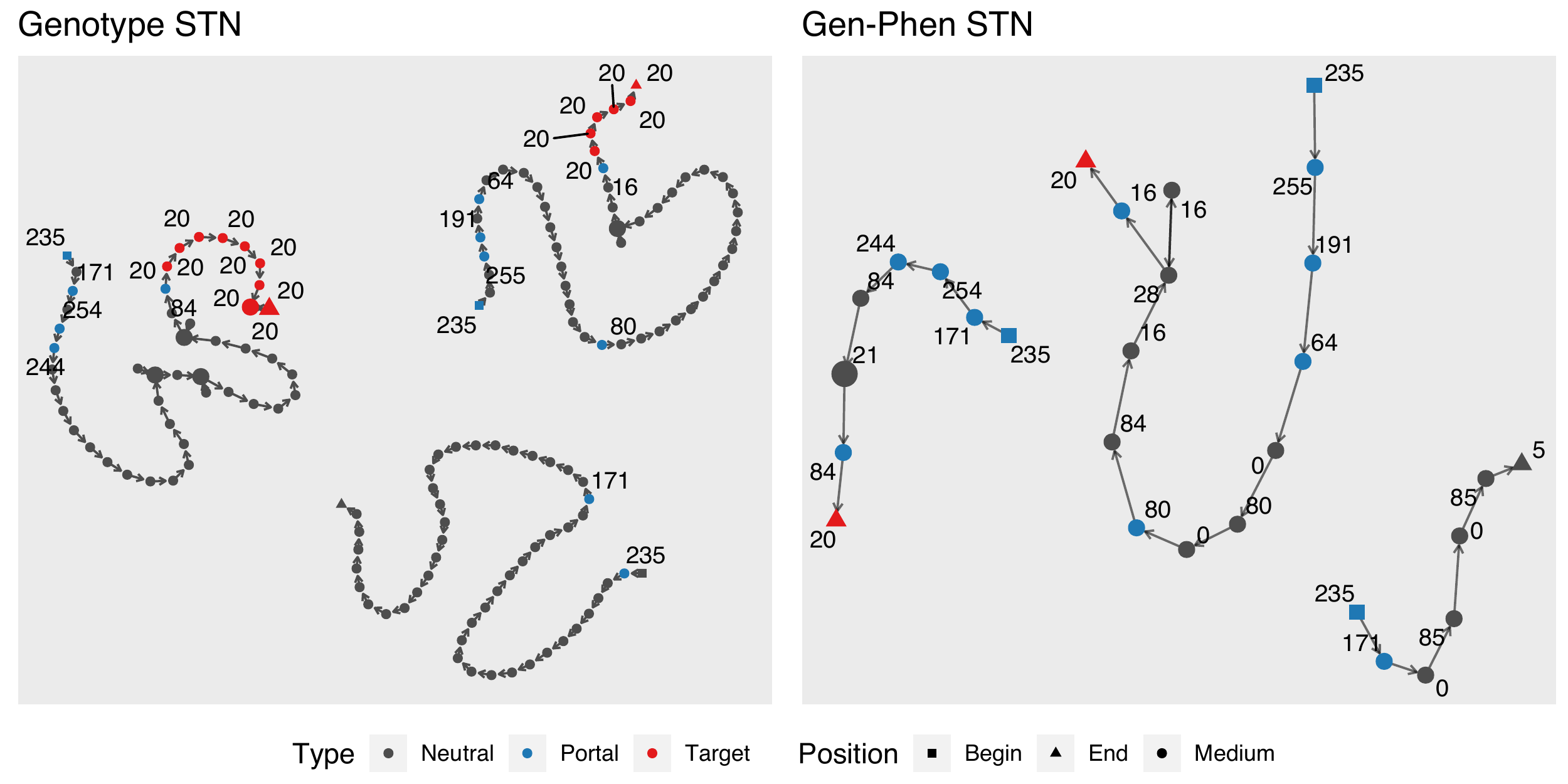}
\caption{Visualisation of the Genotype and Genotype-Phenotype STN models for target 20, 
using a force-directed graph layout.}
\label{fig:gen_phen_stn_t20}
\end{figure}

\begin{figure}[t]
\centering
\includegraphics[width=1.0\linewidth]{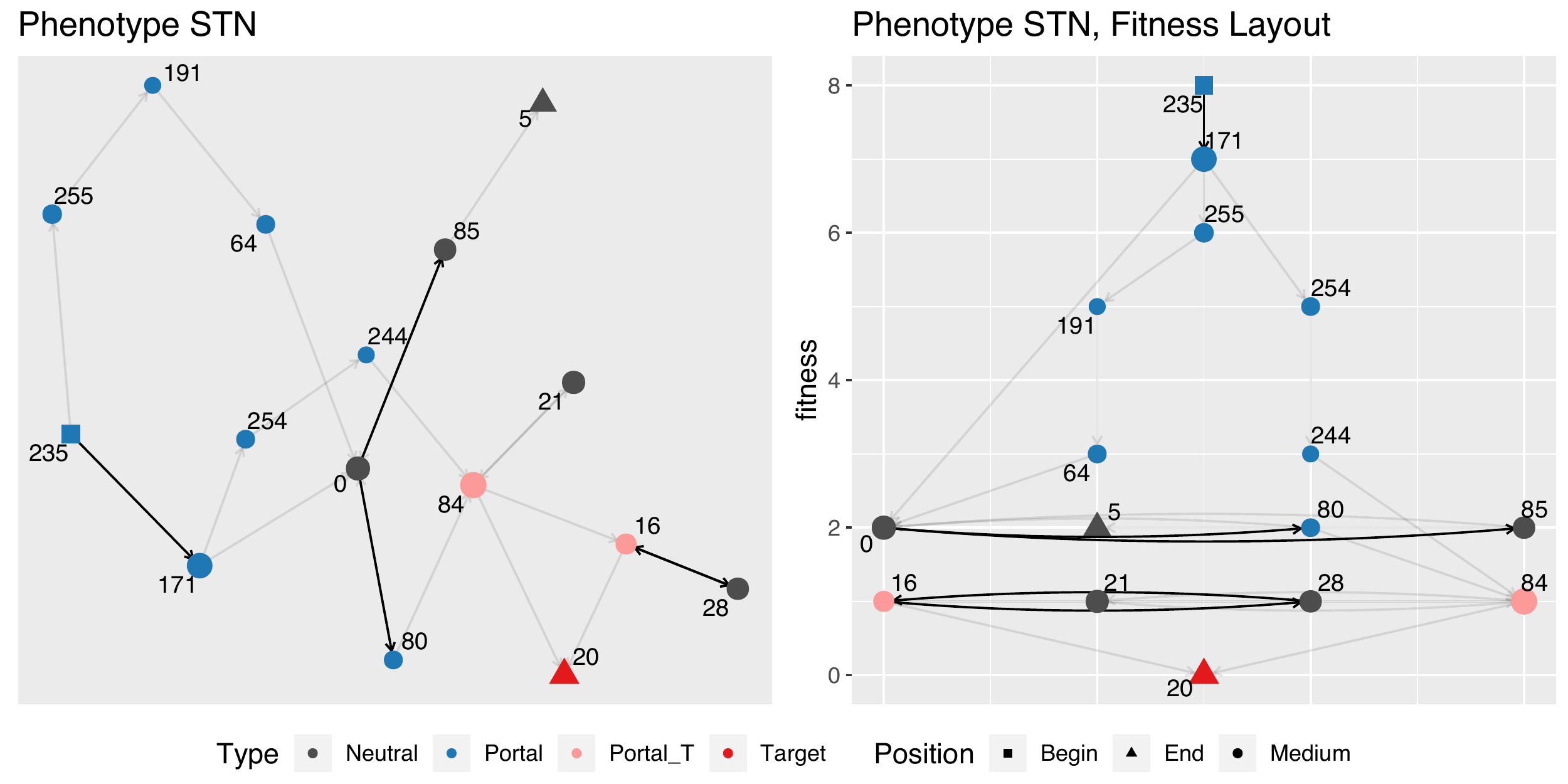}
\caption{Visualisation of the Phenotype STN model for target 20, 
using a force-directed graph layout (left) and a grid layout with fitness in the $y$ coordinate.}
\label{fig:phen_stn_t20}
\end{figure}

To illustrate our proposed STN models, 
we conduct a preliminary experiment using phenotype target 20 (medium difficulty), 
with three runs and 50 steps for the adaptive walks. 
Each run starts from a randomly generated genotype that has phenotype 235. 
Figures~\ref{fig:gen_phen_stn_t20}  and \ref{fig:phen_stn_t20} illustrate the STN models. 
Our STN visualisations use node colors to identify four types of nodes: 
(1)~neutral nodes, whose adjacent outgoing node has the same fitness, 
(2)~portals, which link to a node with improved fitness,  
(3)~target nodes, which have the required phenotype, 
and (4)~(for the phenotype STNs only), we differentiate portal nodes with a direct link to the target.  
The shape of nodes identifies three positions in the search trajectories: 
(1)~begin of trajectories, 
(2)~end of trajectories, 
(3)~intermediate locations in the trajectories. 
Node labels indicate phenotype, 
while node sizes and edge darkness are proportional to their sampling frequency.

On the genotype STN (left plot) in Figure~\ref{fig:gen_phen_stn_t20} 
we can observe the three trajectories corresponding to the three adaptive walks conducted. 
The trajectories are long (remember walks have 50 steps in this experiment) and do not overlap, 
that is, they all visit different genotypes. 
Two of the trajectories reach the target (phenotype 20) while one of them ends in a different phenotype. 
To avoid a cluttered image, the genotype STN plot shows the node labels for portal and target nodes only. 
Notice the long chains of neutral nodes (dark gray) before finding a portal (blue nodes) to improving fitness, 
also several different genotypes in red correspond to the target phenotype.   
The genotype-phenotype STN (right plot) shows shorter trajectories as expected 
as nodes now represent sub-networks joining connected genotypes with the same phenotype. 
Still, the three trajectories do not have overlapping nodes, 
indicating that the three walks visit different regions of the search space.  
Interestingly, there are still long chains of neutral moves (dark gray nodes), 
especially visible in the middle trajectory; 
we can see how the trajectory enters in an out phenotypes 0 and 80, 
before finding a portal to phenotype 84.

\begin{figure}[h!]
\begin{center}
\includegraphics[width=0.95\textwidth]{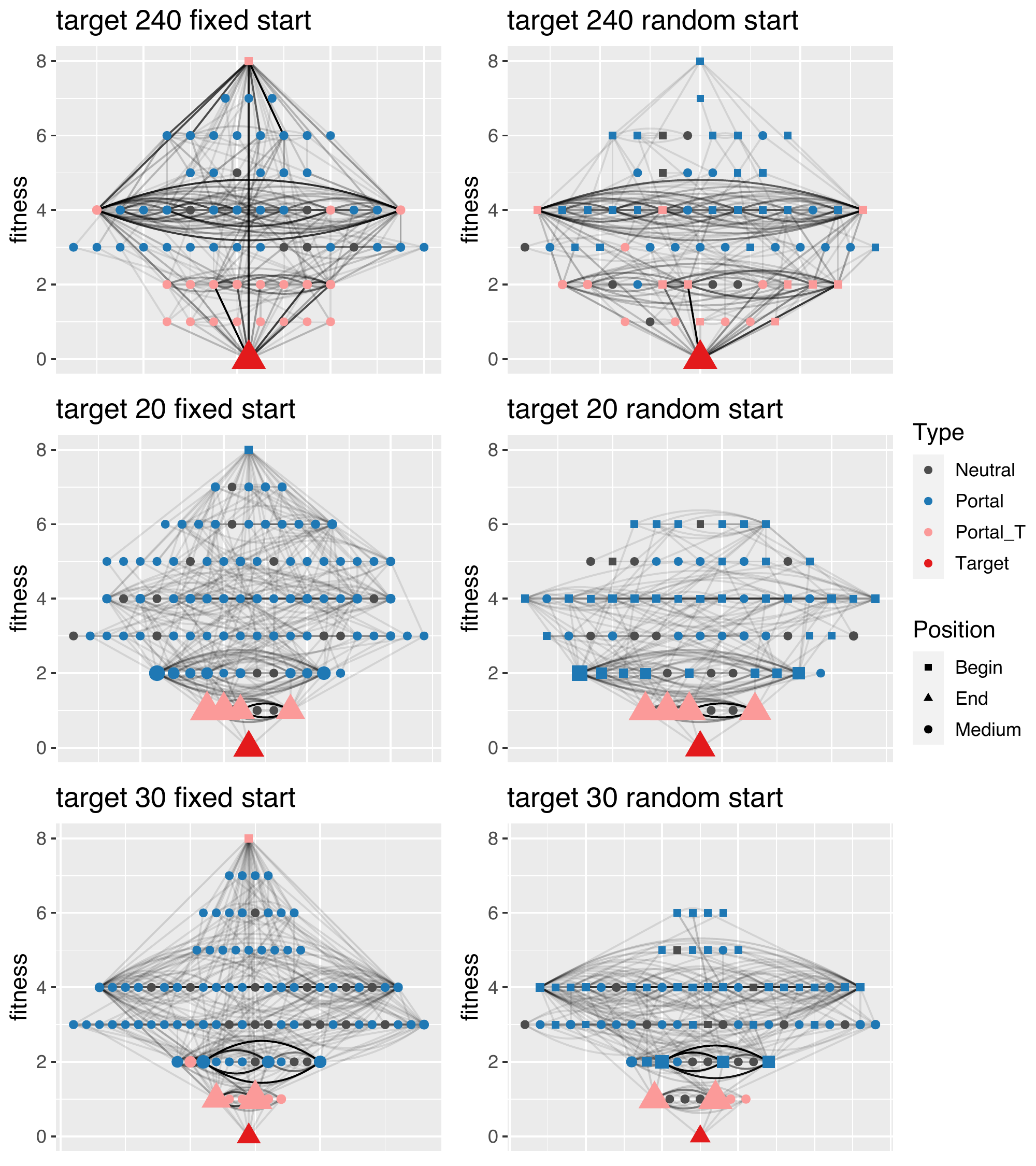}
\caption{Phenotype STNs when searching for target phenotypes of different difficulty 
(240: easy; 20: medium; 30: hard).  
The plots aggregate 100 trajectories, 
which start from either a fixed phenotype (left plots) or a random phenotype (right plots). 
The target node (red triangle at the bottom of each graph) is reached via different search pathways.  
The size of nodes and the darkness of edges indicate their sampling frequency. 
Arrow heads and node labels are omitted to simplify the images.}
\label{fig:stn_all_targets}
\end{center}
\end{figure}

A key aspect of network visualisation is the graph-layout, 
which accounts for the positions of nodes in the 2D Euclidean space. 
Graphs are mathematical objects, they do not have a unique visual representation. 
Many graph-layout algorithms have been proposed. 
{\em Force-directed} layout algorithms~\cite{Fruchterman:1991}, 
are based on physical analogies defining attracting and repelling forces among nodes. 
They strive to satisfy generally accepted aesthetic criteria such as an even distribution of nodes on the plane, 
minimizing edge crossings, and keeping a similar edge lengths.  
We use force-directed layouts for visualizing the STNs models in Figures~\ref{fig:gen_phen_stn_t20} 
and \ref{fig:phen_stn_t20} (left plot). 
For the phenotype STNs, we also introduce a layout that takes advantage of the fitness values. 
The idea is to use the fitness values as the nodes' $y$ coordinates, 
while the $x$ coordinates are placed as a simple grid 
(Figures~\ref{fig:phen_stn_t20} and \ref{fig:stn_all_targets}), 
where nodes are centered according to the the number of nodes per fitness level. 
These plots allow us to appreciate the progression of the search trajectories towards lower (better) fitness values, 
as well as the amount of neutrality present in the search space. 

Our graph visualisations were produced using the \textsf{igraph} and \textsf{ggraph} packages 
of the \textsf{R} programming language.  
The phenotype STN model seen in Figure~\ref{fig:phen_stn_t20} is more compact, 
having fewer nodes and edges as compared the the genotype and genotype-phenotype STN models. 
Most importantly, the phenotype STN model shows search overlaps across the different trajectories. 
That is, there are nodes that have more than one incoming edges, they are hubs, indicating locations that attract the search process.  
For the remaining of our analyses, 
we decided to use the phenotype STN model with the fitness-based graph layout. 
We argue that this combination has a greater potential to reveal interesting aspects of the search dynamic, 
as it allows to observe locations of the search space where the process converges. 
The other models are however interesting to appreciate additional details.

\subsection{Comparing three targets with increasing difficulty}


As described in Section~\ref{sec:sampling}, 
for adaptive walks we set three targets with increasing difficulties (240: easy; 20: medium; 30: hard) 
and two search scenarios (fixed start and random start).
Figure~\ref{fig:stn_all_targets} shows the phenotype STNs for the six configurations. 
The nodes and edges are as defined in Section~\ref{sec:stns}, 
the fitness-based graph layout is used, 
and the arrow heads as well as the node labels are omitted to keep the images less cluttered. 
Notice that the edges are either descending to lower fitness levels or neutral at the same fitness levels. 
The neutral edges are visualised as curves
where the edges above point to the left and the edges below point to the right.

Search proceeds through hops, 
indicated by links of different darkness symbolizing 
how often they were traversed during the sampling process. 
The target node (red triangle at the bottom of each graph) is reached via different search pathways. 
The size of nodes is proportional to how many times it was visited during the adaptive walks, 
so large nodes represent locations that attract the search process. 
For the medium and hard targets (phenotypes 20 and 30),  
many search trajectories do not reach the target, 
they end at phenotypes close in Hamming distance to the target 
(visualised by large pink triangles at fitness level 1).  

It is interesting to observe that the varying size of nodes is more pronounced for more difficult targets.  
Clearly, the landscape becomes more difficult to navigate closer to a difficult target. 
The number of one-step mutant neighbors to a target is smaller for more difficult target nodes, 
signalling that transitions have become more heterogeneous at those levels. 

We can see that most phenotypes are portals (blue nodes) 
offering the possibility of jumping to a lower level fitness,  
but clearly, many neutral moves happen on the way to the target, at each fitness level. 

The graph layout reflects the structure of the search space 
- most phenotypes are located at around half Hamming distance to the target, 
that is at fitness levels 3, 4 and 5. 
The square node at the top of the left plots reflects the fact that 
these trajectories start with a fixed phenotype, 
while the bottom triangle in all plots reflects that the target was found.  
Notice that the size of the red triangle is the largest for the easy target 240, 
and then gradually decreases in size for targets 20 and 30.  
This makes sense as the harder the target the less frequently it is reached by the search process within 2,000 steps.

\section{Discussion}

We now want to connect the observations from these visualisations 
with the complexity considerations mentioned in Section~\ref{sec:complexity}. 
We focus first on target 30, shown in the last row of Figure~\ref{fig:stn_all_targets}. 


\begin{figure}[t]
\begin{center}
\includegraphics[width=\textwidth]{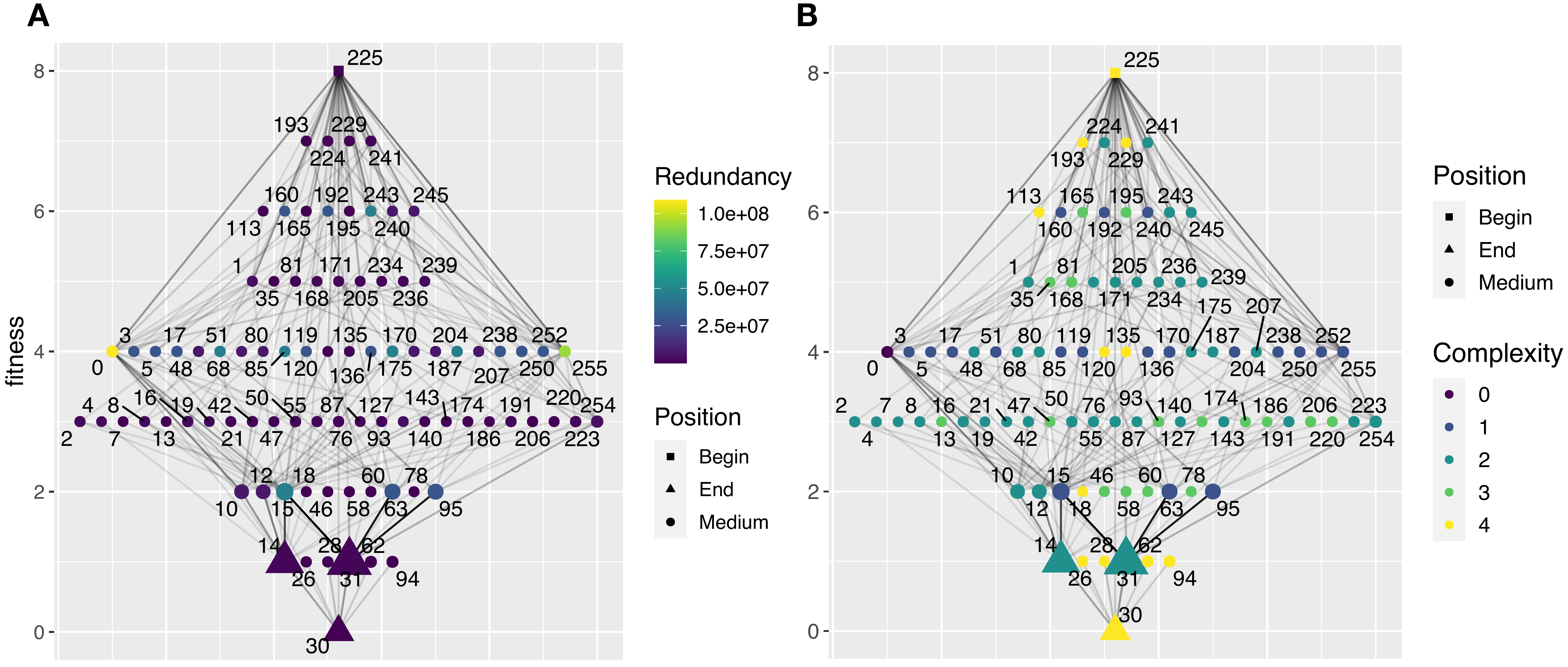}
\caption{Fitness changes for target 30 fixed start runs, 
with phenotype redundancy (A) and Kolmogorov complexity (B) marked by color.
Only non-neutral mutations that improve fitness are shown as edges.
Nodes are labeled with numbers representing their phenotypes.
Shapes stand for different positions in a search.
Size of a node indicates frequency of visit.} 
\label{fig:target30}
\end{center}
\end{figure}

We are interested in more details of the search, 
especially given the heterogeneity at the end of the search, close to the target. 
Table~\ref{tab:1m-target30} shows the phenotypes closest to the target (one-bit mutants) found by the searchers. 
In Figure~\ref{fig:target30}, 
we look at the frequency of fitness-changing jumps from phenotype to phenotype in fixed start runs. 
We label each node with the phenotype it stands for,
with a side by side comparison of nodes color-marked by redundancy (A) and complexity (B). 
We also remove the neutral edges to declutter the images.

We can see that their size strongly correlates with both their redundancy (positively) 
and with their complexity (negatively). 
Recall that larger node size indicates more frequent visits by searchers 
in the process of looking for the target. 
The exponential relationship indicated by equation~\ref{eq:prob} seems to bear out: 
Searchers are much more likely to pass through low complexity/high redundancy nodes 
-- in this case phenotypes 14 and 31 -- 
than through the other one-bit mutant neighbors found, 26, 28, 62 or 94. 

\begin{table}[t]
\begin{center}
\caption{Mutant phenotypes with one-bit distance from target phenotype 30. We characterize the size based on Fig.~\ref{fig:stn_all_targets}, last row and list their redundancy and K-complexity.}
\label{tab:1m-target30}
\begin{tabular}{|c|c|c|c|c|}
\hline
Phenotype &  Node Size  & Node Size & Redundancy &   
Kolmogorov \\
Number &  (fixed start) & (random start) &  & Complexity\\
\hline
14 &  Large & Large & $1.3 \times 10^6$ & 2 \\
22 & N/A & Small & 0  & 8 \\
26 &  Small & Small & $1.2  \times 10^3$ & 4 \\
28 & Small & Small & $1.2  \times 10^3$ & 4 \\
31 & Large & Large & $1.4 \times  10^6$ & 2 \\
62 & Small & Small & $2.9  \times 10^3$ & 4 \\
94 & Small & Small & $2.9  \times 10^3$ & 4 \\
\hline
\end{tabular}
\end{center}
\end{table}

We can extend this analysis to the mutants of the target with two-bit phenotypic distances as 
Figure~\ref{fig:target30} shows all fitness changing moves of searchers for target 30. 
If we focus on two-bit mutants (fitness 2), 
we can see that most transitions happen from the highly redundant phenotypes, first 15, 
followed by transitions from 63 and 95. 
Most of them transition to the highly redundant phenotypes 14 and 31 on fitness level 1.
We can examine in more detail the redundancy/complexity of two-bit mutants. 
Due to the quick combinatorial explosion, 
we have done that in Table~\ref{tab:2m-target30} 
only for the two most representative nodes of fitness distance 1, phenotypes 14 and 31. 
As most redundant stand out the nodes 15, 63 and 95
which are providing the most avenues to a better fitness, 
to a somewhat lesser extent also phenotypes 10 and 12. 

\begin{table}[t]
\begin{center}
\caption{Selected two-bit mutants of phenotype 30: One-bit mutants to the most frequent 1-bit neighbors 14 and 31 of the target node 30.}
\label{tab:2m-target30}
\begin{tabular}{|c|c|c||c|c|c|}
\hline
Phenotype &  Redundancy & Kolmogorov & Phenotype  & Redundancy & Kolmogorov\\
 (to 14) &   & Complexity &  (to 31) &  & Complexity\\
\hline
6 &    $3.0 \times 10^3$ & 4 & 15 & $4.7 \times 10^7$ & 1\\
10 &  $7.1 \times 10^6$ & 2 & 23 & $5.5 \times 10^3$ & 4\\
12 &   $7.1  \times 10^6$ & 2 & 27 & $1.2 \times 10^4$ & 4\\
15 &  $4.7  \times 10^7$ & 1 & 29 & $1.2 \times 10^4$ & 4\\
46 & $4.6 \times  10^4$ & 3 & 63 & $2.9 \times 10^8$ & 1\\
78 &  $4.6  \times 10^4$ & 3 & 95 & $2.9 \times 10^8$ & 1\\
142 &  $2.0  \times 10^3$ & 4 & 159 & $3.1 \times 10^3$ & 4\\
\hline
\end{tabular}
\end{center}
\end{table}

Thus we can explain the dynamics of the search process post-facto 
by looking at the redundancy/complexity of phenotypes in the neighborhood of the target. 
We don't need to know many details of the search, 
except what constitutes the neighborhood of a node, 
to figure out where most searchers will come from. 

Both, redundancy and complexity, require -- of course -- measurements to allow this explanation. 
While they are different (redundancy can be measured for all nodes in parallel), 
it might be argued that one has to have a clear picture of the fitness landscape for this analysis. 
This is correct for a measurement of redundancy, 
but the relationship with complexity is not based on anything 
other than the structure of the phenotypes themselves. 
Thus, in principle, it can be performed completely separate from the search process. 
Evolution is doing here nothing else than seeking out the most probable pathways to the target.
In other words, we can not only explain the search dynamics post-facto, 
but we can try to predict, at least approximately, a search dynamics before it happens. 
This is in line with what other research groups have found 
in their respective systems~\cite{barrick2020,dingle2023,lobkovsky2011}.

\begin{figure}[t]
\begin{center}
\includegraphics[width=.63\textwidth]{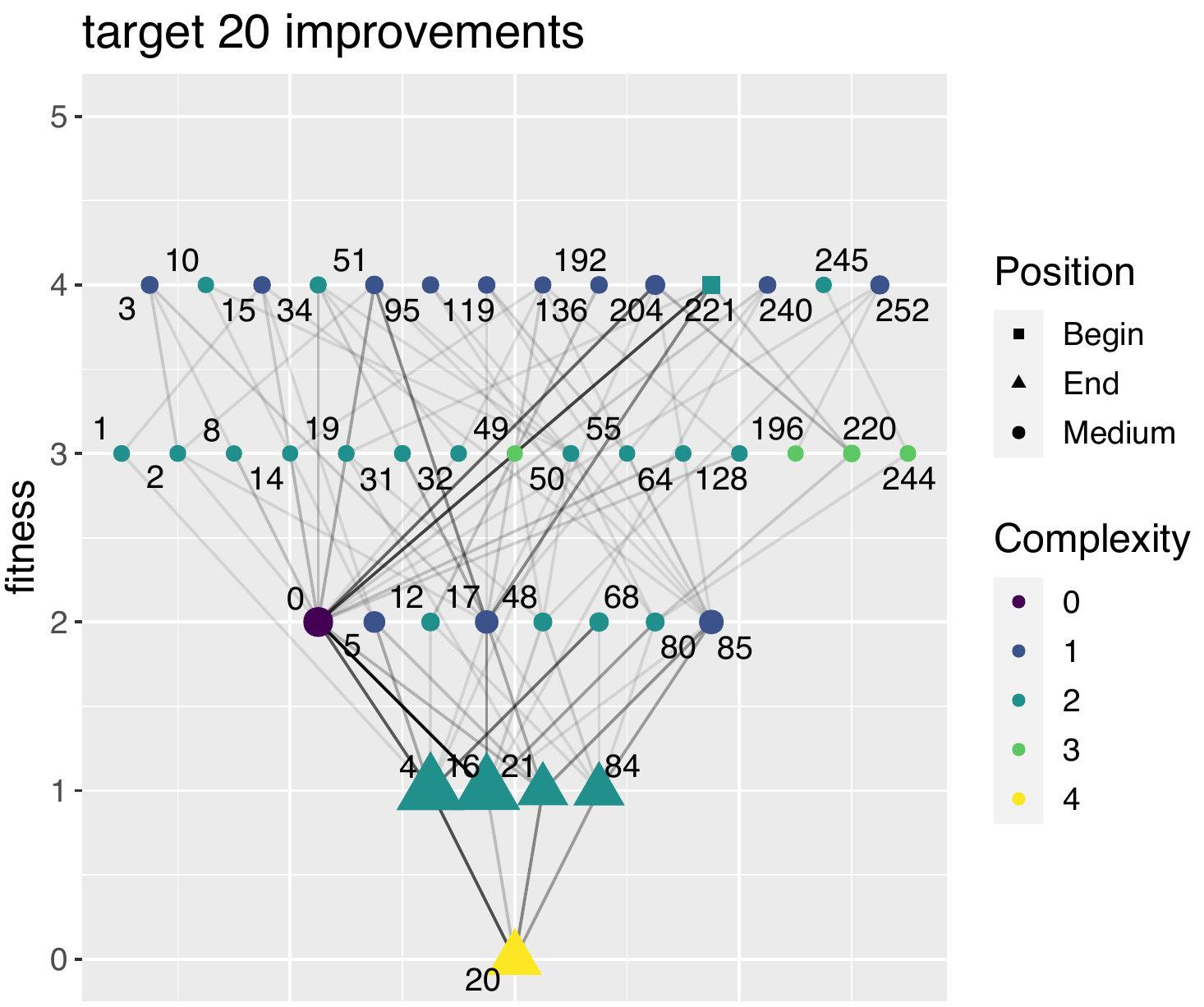}
\caption{Fitness changes for target 20 with fixed starting phenotype 221 (the square). 
Fitness-improving search trajectories for 100 runs with 2,000 steps, 
all going through low-complexity phenotypes to the target from the starting point. 
Phenotype complexity is marked by color.} 
\label{fig:start221_target20}
\end{center} 
\end{figure}

Suppose we start at phenotype 221 to reach target phenotype 20 
(see Figure~\ref{fig:start221_target20}). 
The one-bit neighbors of 221 are 
the set: $\{ 93, 157, 205, 213, 217, 220, 223, 253 \}$. 
Some of those nodes are mutants pointing in the wrong direction of fitness, 
and can be removed from this list because selection would not allow them. 
That leaves us with $\{ 93, 157, 213, 217, 220 \}$. 
However, a brief inspection of the redundancies of these phenotypes
tells us that they are way less redundant ($1.3 \times 10^6$ maximum for 213) 
than the phenotypes of the neutral network, 
which has nodes of redundancies of up to $4.7 \times 10^7$. 
As a result, the nodes $\{ 15, 51, 204, 240 \}$ are the most likely nodes to be accessed 
from 221 on the neutral level and more likely than the one-bit mutations. 
In fact, we can see that only phenotype 220 of the one-bit mutants appears to be accessed. 
But there is one interesting twist here: 
Phenotype 0, being the most likely phenotype in the whole network, 
is a two-bit mutation from 221 in the right direction. 
We can see that it is accessed more frequently than other nodes, 
both directly and indirectly from 221.
Also, phenotype 48, with an redundancy of $7.1 \times 10^6$ is accessed, 
again a two-bit mutation from 221. 
The figure shows that both nodes have lower complexity than 221 
which we know is correlated with their redundancy. 
If we recall Figure~\ref{fig:redun_complexity}A, 
statistics shows that two-bit mutations are actually more frequent than one-bit mutations in this system, 
followed by four-bit mutations. 
It seems that the step size is less of a concern for the searchers than the redundancy of phenotypes!

We note in passing that there are many more pathways to a higher fitness solution 
when not only the rearrangement of instructions is possible 
(as would be the case in a transition from a program with six effective instructions) 
but when also an increase in the number of effective instructions were possible 
(as would be the case in a transition from a program with a smaller number of effective instructions).

Why is there such a strong correlation between phenotypic redundancy and K-complexity? 
This is an important question since - as we have seen - 
redundancy has such an influence on the trajectories taken by adaptive searchers in this fitness landscape. 
The answer has to do with the hard length limit in our system, 
which allows a maximum length of programs of six instructions. 
Suppose a phenotype has a K-complexity of 2, 
thus is not using the other 4 instructions theoretically available, 
they are rendered non-effective. 
A brief combinatorial consideration allows us to estimate that 
there are maximally $200^4 = 1.6 \times 10^9$ programs with four neutral instructions, 
assuming all calculation registers are used as a destination. 
This will be an upper limit, of course, as many of those might well not be neutral, 
either by virtue of their order or because of their internal composition. 
Nevertheless, it is a huge number of neutral variations of the same program. 
Compare that to an individual with five out of the six instructions being effective. 
We only have one instruction left that can be neutral, 
leaving a maximum of $200$ neutral variations for this program.~\footnote{Note that multiple programs can contribute to the same phenotype as specified by its behavior.}

In summary, the reason why K-complexity is negatively correlated with redundancy of programs 
and thus phenotypes is the combinatorics in the neutral space! 
While this presupposes a hard limit on the total length of programs 
(effective plus non-effective code), 
we expect that a soft limit can allow similar effects to play out, like in RNA. 
It will probably not be as clearly visible, 
but should still be expected to emerge in such systems. 
These considerations are not restricted to the particular system we have examined here: 
The combinatorics of neutral spaces determines the redundancy of phenotypes 
and thus to a substantial degree the search trajectories in length-changing evolutionary systems in general.

\bibliographystyle{splncs04}
\bibliography{EuroGP2023}
\end{document}